\documentclass[preprint2]{aastex}
\usepackage{graphicx}

\shorttitle{VLBA fast transients experiment}
\shortauthors{Wayth et al.}

\begin{document}


\title{V-FASTR: The VLBA Fast Radio Transients Experiment}

\author{Randall B. Wayth\altaffilmark{1},
        Walter F. Brisken\altaffilmark{2},
        Adam T. Deller\altaffilmark{2,3,5},
        Walid A. Majid\altaffilmark{4},
        David R. Thompson\altaffilmark{4},
        Steven J. Tingay\altaffilmark{1} and
        Kiri L. Wagstaff\altaffilmark{4}
}
\email{randall.wayth@icrar.org}

\altaffiltext{1}{International Centre for Radio Astronomy Research, Curtin University. GPO Box U1987, Perth WA, 6845. Australia}
\altaffiltext{2}{NRAO, PO Box O, Socorro, NM 87801, USA}
\altaffiltext{3}{601 Campbell Hall, University of California at Berkeley, Berkeley, CA, 94720, USA}
\altaffiltext{4}{Jet Propulsion Laboratory, California Institute of Technology, 4800 Oak Grove Drive, Pasadena, CA 91109, USA}
\altaffiltext{5}{Adam Deller is a Jansky Fellow of the National Radio Astronomy Observatory}

\begin{abstract}
Recent discoveries of dispersed, non-periodic impulsive radio signals with single-dish radio telescopes have sparked significant interest in exploring the relatively uncharted space of fast transient radio signals. Here we describe V-FASTR, an experiment to perform a blind search for fast transient radio signals using the Very Long Baseline Array (VLBA). The experiment runs entirely in a commensal mode, alongside normal VLBA observations and operations. It is made possible by the features and flexibility of the DiFX software correlator that is used to process VLBA data. Using the VLBA for this type of experiment offers significant advantages over single-dish experiments, including a larger field of view, the ability to easily distinguish local radio-frequency interference from real signals and the possibility to localize detected events on the sky to milliarcsecond accuracy. We describe our software pipeline, which accepts short integration ($\sim\mathrm{ms}$) spectrometer data from each antenna in real time during correlation and performs an incoherent dedispersion separately for each antenna, over a range of trial dispersion measures. The dedispersed data are processed by a sophisticated detector and candidate events are recorded. At the end of the correlation, small snippets of the raw data at the time of the events are stored for further analysis. We present the results of our event detection pipeline from some test observations of the pulsars B0329+54 and B0531+21 (the Crab pulsar).

\end{abstract}


\section{Introduction}
\label{sec:intro}
Recently \citet{2007Sci...318..777L} reported the detection of a single, short-duration ($\sim$5~ms), powerful, highly dispersed burst of radio emission originating from an unknown object.
This “fast transient” event, captured by virtue of a search through existing pulsar survey data from the Parkes radio
telescope, has generated significant interest because it was interpreted to be extragalactic in origin.
While this event was proposed to be astronomical in nature by \citet{2007Sci...318..777L}, recent work by
\citet{2011ApJ...727...18B} has revealed an additional 16 events in the Parkes data with characteristics that indicates a non-astronomical but otherwise unknown origin.
Important differences exist between the event reported by \citet{2007Sci...318..777L} and those reported by \citet{2011ApJ...727...18B}, but the new work casts doubt on the astronomical origin of the \citet{2007Sci...318..777L} event, prompting the need for further investigation into the existence and characterization of these types of events. The existence of extragalactic pulses would be valuable scientific probe into both the physics of extreme objects and the properties of the inter galactic medium.

The origin of the \citet{2007Sci...318..777L} event is yet to be determined, but it is possible that short-timescale, highly dispersed bursts of radio emission could originate from a variety of known and speculative compact objects in the universe
such as neutron stars, annihilating black holes, gravitational wave events or even extraterrestrial civilizations \citep[see][and references therein]{2010PASA...27..272M,2010CRAFT}.
Specially designed experiments are required to explore these possibilities.
In this paper we describe one such experiment that runs in a fully commensal fashion on the National Radio Astronomy Observatory's (NRAO) Very Long Baseline Array \citep[VLBA; ][]{1994IEEEP..82..658N}.
With transient radio astronomy of increasing interest for next generation radio telescopes such as the Australian SKA Pathfinder \citep[ASKAP; ][]{2008ExA....22..151J,2007PASA...24..174J}, the Murchison Widefield Array \citep[MWA; ][]{2009IEEEP..97.1497L}, the Allen Telescope Array \citep[ATA; ][]{2009IEEEP..97.1438W}, LOFAR \citep{2009ASPC..407..318H} and the Square Kilometre Array (SKA) itself \citep{2009IEEEP..97.1482D}, experiments such as described here are important trailblazer activities, to determine if the science from fast transient radio astronomy is likely to be a design driver for next generation instruments.  In an unexplored region of signal parameter space, simple trailblazer experiments can be highly informative.

Traditionally, imaging radio telescopes have been insensitive to such short-timescale events, since data averaging times on the order of seconds are generally used, with no allowance for the effects of signal dispersion.  The exceptions are observations for the detection of pulsars that use averaging times on the order of ms or smaller.
Also, pulsar observations are typically performed by large single dishes, instruments that have poor angular resolution, a narrow field of view (FOV) and can be severely affected by radio frequency interference (RFI) from the terrestrial or near-Earth activities of humans.
A narrow FOV is also mismatched to detecting rare events such as that seen by \citet{2007Sci...318..777L}, which are assumed to be isotropically distributed over the sky.
Despite this, efforts dedicated to reprocessing pulsar data for non-periodic transient signals have been very fruitful in discovering a new class of neutron stars \citep{2006Natur.439..817M} as well as the bursts seen by \citet{2007Sci...318..777L} and \citet{2011ApJ...727...18B}.

With this in mind, an experiment to detect astronomical bursts should include:
(1) the capability to produce high time and frequency resolution data,
(2) a distributed set of telescopes that are only affected by the RFI local to each telescope and not common RFI signals,
(3) a broad frequency coverage,
(4) good sky coverage through FOV, and
(5) the ability to account for large and unknown dispersion measures (DMs) over a variety of observing frequencies.
With these attributes an experiment can detect bursts, discriminate against RFI by searching for signals common to all telescopes, and localize the burst on the sky with enough angular resolution to identify the origin of the burst.

Identifying the origin of any burst is critical to understanding the physics of the emission process and nature of the emitting object, since distances can then be determined (at least in the case of an extragalactic source, which can presumably be associated with a galaxy) and observational parameters such as the DM can be used to derive useful information regarding the average line-of-sight electron density, for example.

In this paper we describe a new experiment, designed for robust detection and precise localization of fast transient radio signals.
The experiment couples an instrument based on distributed radio telescopes, the VLBA, with a flexible and powerful software correlator
that generates time-aligned high time and frequency resolution data \citep[DiFX;][]{2007PASP..119..318D,2011PASP..123..275D}, and machine learning algorithms \citep{Thompson2010} for the detection of short timescale dispersed bursts of radio emission. The experiment, by its nature, allows the possibility of localizing the position of any astronomical burst with milliarcsecond angular resolution and sufficient signal-to-noise ratio (S/N). Furthermore, the experiment runs in a fully commensal fashion with regular VLBA operations and therefore does not require additional resources to execute, resulting in a blind survey of the sky.

\section{System Description}
\label{sec:sys_descr}

\subsection{Basic Requirements and Constraints}
\label{sec:reqs}

The overall aim of V-FASTR is to detect and localize short bursts of radio emission. Radio waves are delayed as a function of frequency by interstellar (and intergalactic) plasma, quantified by the well known cold plasma dispersion relation
\begin{equation}
\Delta t = 4.15 \mathrm{ms}  \left( \mathrm{DM} \right)  \left[ \nu_{\mathrm{lo}}^{-2} - \nu_{\mathrm{hi}}^{-2} \right]
\end{equation}
where the DM is the column density of electrons along the line of sight to the source (units $\mathrm{cm}^{-3}\mathrm{pc}$), and $\nu_{\mathrm{hi}}$ and $\nu_{\mathrm{lo}}$ are the highest and lowest frequencies, respectively, in the bandwidth of interest in GHz. Optimally detecting transient radio signals depends on many factors, as discussed in detail by \citet{2003ApJ...596.1142C}.
The key requirements for a total-power (or incoherent) system such as V-FASTR are to have suitably fine time and frequency resolution spectrometer data, to be able to dedisperse the data and to be able to detect events in the dedispersed data. For a system that has many antennas such as the VLBA, it is also highly advantageous to dedisperse the data from each telescope independently and combine the data only at the detection stage.

Regular VLBA observations involve recording raw baseband data at each telescope on portable disk packs.
The disks are shipped to the NRAO's Science Operations Center (SOC) in Socorro, NM for correlation.
An observation is correlated by loading the appropriate disk packs at the SOC where the data can be read by
the DiFX software correlator \citep{2007PASP..119..318D}. Shortly after the correlation job has completed, the disks
are unloaded and shipped back to the telescopes for re-use.

The VLBA uses the DiFX software correlator \citep{2007PASP..119..318D} for regular correlation tasks. In addition, the new DiFX 2.0
correlator \citep{2011PASP..123..275D} is available for test and special-purpose observations. DiFX 2.0 is currently being phased in and is expected to become the default correlator in mid 2011.
DiFX 2.0 has many useful features, of which the most important for V-FASTR is the ability to output short integration spectrometer data for each antenna while the correlation job progresses. Because DiFX is an FX correlator, the spectrometer data can be generated with minimal extra cost as part of the ``F'' (Fourier transform) stage of correlation. The properties of the spectrometer data, such as time and frequency resolution, are configurable. We aim for approximately 1~ms time resolution for observing frequencies around 1.4~GHz, which is a good compromise between signal detectability and ensuring that there is no impact on correlator operations.
The spectrometer data stream generated during correlation forms the raw data for the V-FASTR processing pipeline. DiFX can also calculate the spectral kurtosis \citep{2007PASP..119..805N} of data on short timescales which is a useful tool for system diagnostics and RFI detection.

The spectrometer and kurtosis data are distributed via the flexible DiFX Message system, which uses network multicasting to allow one or more clients to receive the data. The underlying network protocol of DiFX Message is UDP (User Datagram Protocol), which was designed for high-throughput data streams for which synchronization is unnecessary and the occasional lost packet is acceptable.
Additionally, DiFX is an unclocked, distributed application that runs over many compute nodes. The nodes are not required to operate in strict time synchronization, so the data from different compute nodes received at the same time can have different time tags.

During processing, DiFX automatically time aligns the data streams to the observation phase center. Transient signals may be received from anywhere in the primary beam, however, which results in an unmodeled extra delay in each data stream that depends on the antenna location and the location of the signal within the primary beam.
The extra delay depends on the source offset relative to the phase center and the antenna location, with minor corrections due to special and general relativistic effects \citep{1998RvMP...70.1393S}.
We can estimate the maximum unmodeled delay by assuming the worst case scenario at 1.4~GHz where $D\sim 6000$~km and the angular offset from the beam center is $\sim 0.3^{\circ}$. In that case, the change in propagation distance is $\sim 3 \times 10^4$~m, so the extra delay is $\sim 0.1$~ms. For $\sim1$~ms integration times this is negligible, however this delay is explicitly handled during follow-up of candidate events as discussed in Section \ref{sec:follow-up}.

An additional complication from using spectrometer data is that many radio telescopes include a noise
calibration signal that is switched into the radio-frequency (RF) stream at the antenna. For the VLBA, each antenna has a
calibration noise source signal switched in with frequency 80 Hz and 50\% duty cycle. The calibration signal
causes the total power across the bandwidth to change by approximately 10\%, so is a clear periodic signal when
the power is summed over the band. This signal must be removed from the data before dedispersion and detection.

V-FASTR aims to have no impact on regular VLBA operations. However, to post-process candidate events
a small section of the raw baseband data around the time of the event
must be copied before the disk packs are unloaded. (For technical reasons, data cannot be copied while a correlation
job is running, only afterwards when the disks are idle.) This requires the transient data processing to be performed in
real time during a correlation job and for the small chunks of raw data to be copied immediately after the job has ended,
before disks are unloaded.

\subsection{Data Processing Pipeline}
\label{sec:pipeline}

\begin{figure}
 \plotone{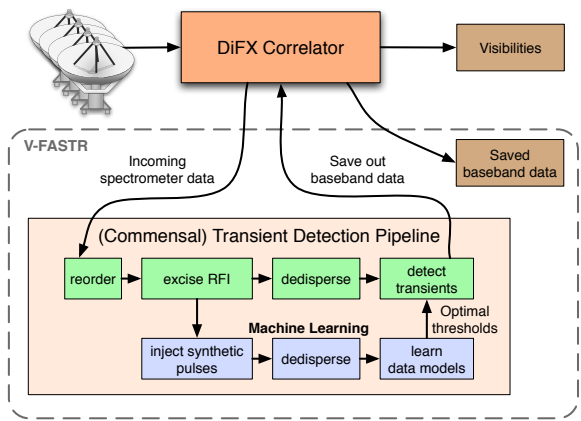}
 \caption{V-FASTR real-time pipeline.}
 \label{fig:pipeline}
\end{figure}

These operational and technical constraints lead to a real-time processing pipeline as shown in Figure \ref{fig:pipeline}.
A transient daemon process (not shown) listens for messages from DiFX signaling the start of a new correlation job.
When a suitable new job is detected, the daemon starts the dispatcher.
Each job is broken into one or more scans, where a scan translates operationally to an unbroken track on a source
with a constant instrument configuration.
Between scans there may be many seconds of dead time (e.g., if a telescope
is slewing), or a configuration change (e.g., a change of frequency), hence the data streams from each scan
must be kept separate.
The dispatcher creates one instance of the pipeline for each scan in the job and funnels the
spectrometer and kurtosis data to the appropriate pipeline by examining the scan index in the data packet headers.

\begin{figure*}
 \centering
 \includegraphics[scale=0.75]{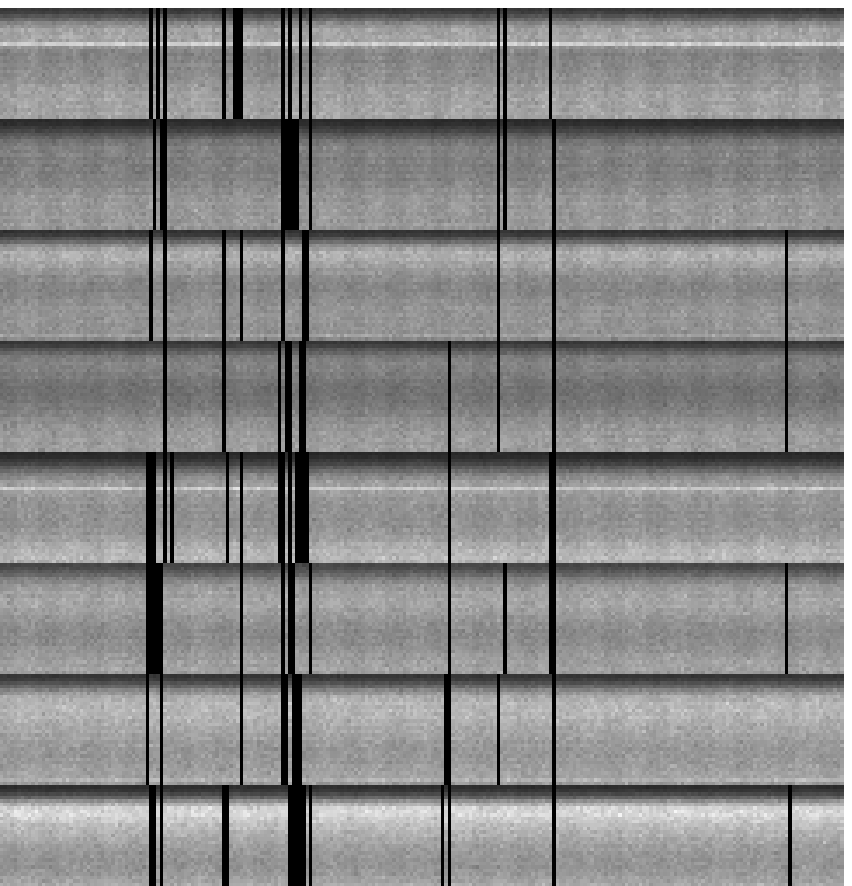}
 \includegraphics[scale=0.75]{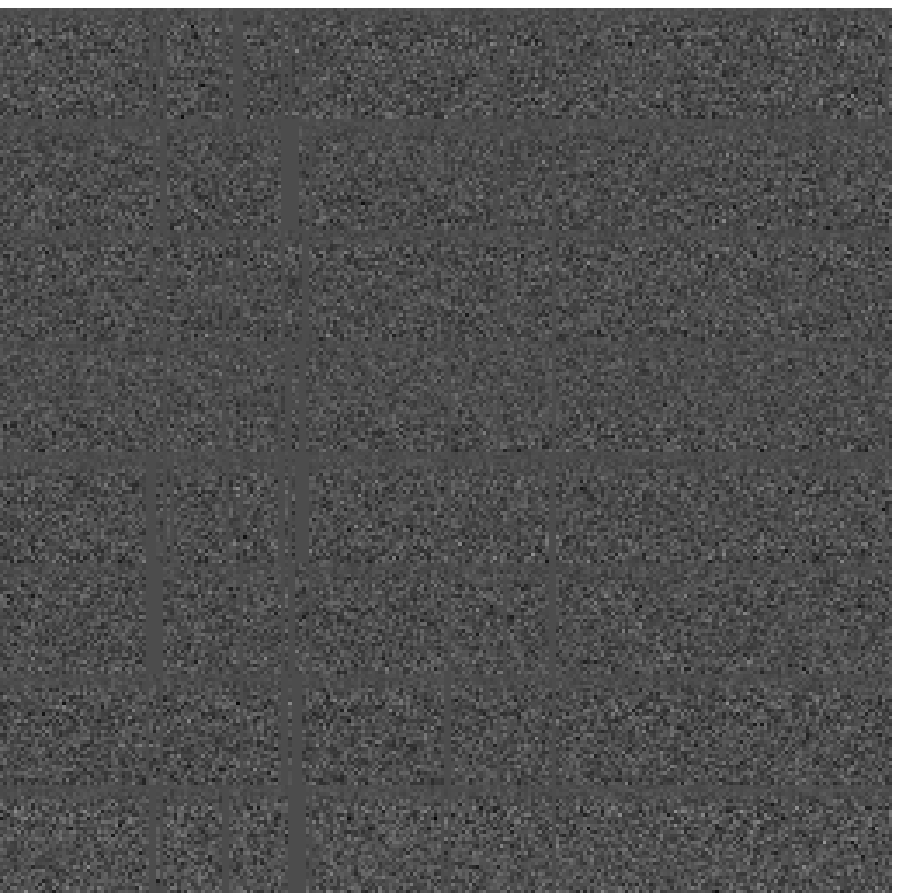}
 \caption{Example raw (left) and post-reorder (right) data. Time runs left to right and frequency increases upwards.
          Each pixel in time is approximately 1~ms. In this example, the bandwidth
          was broken into 8 bands of 32 channels each. The dark edge of each band is due to the roll-off of the
          analog filters. Missing data are obvious as black vertical strips in the raw image. Also visible is the
          calibration signal which runs vertically throughout all bands. After the reorder the output is a
          time-ordered zero-mean data stream. There are no transient signals in this example, so the output is noise-like.}
 \label{fig:reorder_example}
\end{figure*}

The first step in the pipeline is the reorder process, which corrects for missing data, out-of-order
arrival times and noise calibration signals. This is achieved by buffering several seconds of data and forming
a running median of the power in each channel, both for when the calibration signal is turned on and for when it is turned off.
The phase of the calibration signal is locked to the clock at each antenna, so it is possible to predict the fraction of calibration
signal that is present in each data packet by using the DiFX delay model. Incoming data packets are sorted into
the buffer and median subtracted for the appropriate fraction of calibration signal that they contain before they are sent on. The output data are thus zero mean. All missing data are zeroed, and thus do not create any artifacts in the (zero mean) data stream.
Median subtracting the data stream so that it is zero mean also has advantages for the dedispersion stage to follow.
Figure \ref{fig:reorder_example} shows an example of a small section of filterbank data before and after the reorder process.
There is one such stream of filterbank data from each antenna and the streams are not combined until the detection step.

Also as part of the reorder process, the pipeline performs optional automatic RFI excision.
The VLBA antennas are vulnerable to RFI, especially at low frequencies between
0.2 and 2.4~GHz.\footnote{See \url{http://www.vlba.nrao.edu/astro/rfi/}}
The RFI is generally narrow band but can have significantly different temporal characteristics such as being bursty
or slowly varying over a second or longer. Despite being mostly narrow band, strong RFI will affect power over the entire
frequency band by skewing the sampler statistics and forcing automatic gain control systems to adjust the gain,
causing the total power in the band to change with time.

The power in each spectrometer channel is the average of hundreds or thousands of individual FFTs, hence we expect the power
in a channel to be very close to normally distributed around some mean power.
In addition, because the telescope data capture systems automatically adjust the gain to achieve ideal statistics with 2 bit samples, we know in advance that the total power in each frequency band should be almost the same.
We can therefore apply some simple heuristics to identify and excise RFI:
\begin{itemize}
\item if the median power in a channel over a short time ($\sim1$ s) is much higher than the typical median power across the band, then it has narrow-band RFI and should be excised for the entire time period;
\item if the power in a channel at any instant is many times the standard deviation of the power over a short time, then it has bursty RFI and should be excised;
\item if the variance in any channel over a short time is much higher than the expected variance, then the power in the band
has been affected by RFI and should be excised for the entire time.
\end{itemize}
In addition, we automatically ignore the ``DC'' channel for all bands because it typically has anomalously high power due to sampler offsets. We use the interquartile range as a robust, yet computationally cheap, estimator for the variance of the power in each channel
\citep{2008AJ....135.1810F}.

The next step in the pipeline performs an incoherent dedispersion across the frequency bands separately for all
antennas. Many trial DMs are used, tailored to the observing frequency, frequency resolution, and integration time, including some negative DMs. Negative DMs are not generated by known astrophysical processes, hence tracking the number of detections for negative DMs will be a useful diagnostic of false positive rates.

We use our own code, DART (a Dedisperser of Autocorrelations for Radio Transients), for the dedispersion step.
DART is specifically designed for data with multiple antennas and multiple frequency bands, such as the VLBA.
Data incoming to DART is ordered by frequency then time, i.e., each chunk of data is a set of channels for
an antenna/band for a specific time. The dedispersion process sums power across channels, with a delay offset for each
channel according to the frequency and DM.
Since the output dedispersed time series for any DM, say for $N$ time steps, is just the sum of $N$ time steps from the
input data with a different delay for each frequency, it is efficient to process the data in the following way:
\begin{enumerate}
\item define a block size as the number of time steps of delay between the highest and lowest frequency for the highest trial DM;
\item create working spaces, one per antenna/band, which is a multiple $M+1$ ($M$ integer greater than 0) of the block size in time;
\item read data into the working spaces. The data are ordered by frequency, then time due to the nature of the incoming data stream;
\item transpose the data in the working space so that they are ordered by time, then frequency;
\item form the output time series for each trial DM by adding $M$ blocks of data from each frequency into the output time series with the appropriate delay;
\item transpose the output block so that time varies most slowly again and write the output as a data stream with implicit dimensions of time (varying most slowly), antenna then DM (varying most quickly);
\item copy the last block to be the first block of the next batch.
\end{enumerate}

The cost of the transpose operations is $O(1)$ for each data sample, whereas each sample is accessed once per trial DM, hence the transpose cost is negligible for many trial DMs. The advantage of this approach is that additions of large vectors of numbers can be highly optimized on modern computing hardware. Because the data are zero mean, it is also trivially easy to incorporate channel flagging by simply not adding a channel into the total. Lastly, since each DM total is independent of the other, it is straightforward to form the totals for each DM in parallel and make full use of modern multi-core compute processors. Different polarizations for the same frequency are also combined at this stage and the headers from data packets
are removed so that the output is a data stream of dedispersed power values.

\subsection{Detection and Sensitivity}
\label{sec:det_sens}

The detection system incorporates novel discriminant algorithms described in detail in a companion paper \citep{Thompson2010}. The input to the detector is a time series of the summed broadband signal from each independent antenna, after dedispersion to each different DM.  The data are a matrix of size [antennas] $\times$ [DMs].  The detector scores each timestep according to the probability that it contains a transient event.  Currently three detectors are implemented: the standard thresholding approach, which simply sums the signal from all antennas, and two adaptive approaches.  The adaptive methods include a robust trimmed estimator that sorts the station's signals by intensity and excises extreme values prior to summing, and an ensemble CDF estimator that combines each station's independent estimate of the probability that the timestep would exceed a random draw from its observation sequence.  The adaptive methods provide resilience to RFI by exploiting the locality of RFI and the statistical correlations of a real pulse event across multiple geographically separated antennas.   For a complete discussion of their sensitivity properties and detection performance, we refer the reader to the companion text \citep{Thompson2010}.

Detection scores are passed as a time series to a subsequent event list generator module that makes a final decision about which segments to archive for further analysis.  The scores from the detection stage are normalized and used to rank order candidate time segments for archiving.  Note that the detector module can consider all antennas and DMs, while the event list generator relies on the single scalar score.  However, the detector is stateless while the event list generator tracks the entire history of observations.  The event list generator thus knows the total disk resources remaining for archiving candidates and can make an informed decision based on the entire history of the job and the correlation time ``budget'' allocated for V-FASTR. Baseband data from all antennas are then archived in prioritized order after the correlation completes.

In addition to the standard stream of dedispersed observation data, the dispatcher generates a parallel data stream into which synthetic pulses are injected at regular intervals.  The S/N and DM of injected pulses varies with the lowest strengths near or just below the detection limit.  This parallel stream undergoes the same dedispersion as the main stream and is processed by a parallel detection stage.  The second stream serves multiple purposes.  First, the SNR of the retrieved pulses can be used to estimate the empirical detection sensitivity.  Such sensitivity estimates incorporate factors that are difficult to model analytically, such as unexpected RFI and the discretization errors introduced during incoherent dedispersion.  They permit principled performance statistics to be gathered for a much broader range of array configurations and detector options.  On-line performance estimates are particularly important for machine-learning-based detection algorithms whose sensitivity may defy closed-form analysis.  In the case of a null result (i.e., no detections), an empirical sensitivity estimate can constrain what events should have been detected, and as a result the actual transient population.

The parallel stream carries other benefits.  Retrieval rates of synthetic pulses provide a measure of data quality that is independently useful to operators of the VLBA as a system performance metric.  In addition, the parallel stream also informs an online learning procedure that dynamically optimizes detector parameters according to the noise character of each observation.  For example, a noisy configuration may require stricter detection thresholds or larger trimming rates in the robust estimation procedure.  The parallel pulse stream offers a constant source of ``training'' examples that can be used to adjust these values and immediately see the effect on retrieval rates.  In practice, the detector accumulates approximately a second of data between retraining sessions.  Retraining optimizes an objective based on the Area Under the Curve metric representing the ability to discriminate pulses from background noise \citep{Thompson2010}.   These retraining results are logged along with the expected performance.

The main V-FASTR pipeline performs an incoherent dedispersion and sum over the frequency channels separately for
each antenna. The VLBA antennas have a system equivalent flux density of approximately 300~Jy at the
$\sim$GHz frequency range that is of most interest to the experiment. Assuming 64~MHz of bandwidth is used with 1~ms
spectrometer integration times, the $1\sigma$ variation in total power should be 1.2~Jy per polarization per antenna.
By incoherently summing the signal over polarizations and all VLBA antennas, an additional improvement of up to
$\sqrt{20}$ can be gained which makes the theoretical $1\sigma$ sensitivity of the experiment 0.27~Jy averaged over 1~ms.
This sensitivity is sufficient to easily detect an event such as that seen by \citet{2007Sci...318..777L}, the brighter tail of pulses seen in 2 of the 11 sources in \citet{2006Natur.439..817M} and one of the ``Peryton'' events seen in \citet{2011ApJ...727...18B}.

\subsection{Candidate Event Follow-up Strategy}
\label{sec:follow-up}

Candidate events that are recorded for follow-up can be confirmed using a series of steps that culminates with visibility data that are suitable to form an image. Ideally, the data should be coherently dedispersed, recorrelated, and imaged at the time of the pulse.
The unknown location of the pulse within the primary beam is an issue, however. For a VLBI array the typical FOV that can be routinely imaged is arcseconds in size due to the delay pattern, thus the location of the pulse within the primary beam should be localized (if possible) to avoid a brute-force search over the primary beam. We will use the following strategy to follow up candidate events.

First, the signal is confirmed and its DM refined by reprocessing the data incoherently, but with finer time resolution. Transient pulses that are not broader than the initial averaging time will have their S/N improved by reprocessing with finer time resolution. In addition, examining the high time resolution data for unmodeled delay between antennas allows the estimation and removal of gross delays (of order hundreds of microseconds or more).  Any such gross delay corrections are saved and added to the delays derived from calibration in subsequent processing to obtain position offsets for the source. The start time and duration of the event are refined for use in the following step.

Next, the data are correlated at high frequency resolution to provide a wide delay window to encompass the delay pattern across the antennas in the array. The visibilities are then dedispersed using the refined DM estimate and formed into a standard FITS file. These data are loaded into AIPS and a priori calibration is performed.  At a minimum, this consists of amplitude corrections based on the measured system temperature values to convert the flux scale into Janskys.  Delay and phase corrections from any calibrators present in the observation can also be applied if available.  VLBA calibration data have no proprietary period and can be obtained using existing VLBA procedures, but this process requires some time and the calibrator data may not be available during this initial imaging search.  Without application of delay solutions from an external calibrator, small clock errors at the VLBA stations will not be corrected and will contribute to errors in the delays measured for the transient source.  However, these delays are typically small (of order 10~ns, similar to our delay solution precision) and can always be corrected at a later reprocessing once the calibrator data have been obtained.

After initial calibration, the AIPS task FRING is used to solve for the antenna-based delays to the source, using the cross-correlation data.  These delays are used along with the visibility $uvw$ coordinates at the time of the event to solve for a position offset for the source.  At this time, this position-solving step assumes an astronomical source -- a poor fit to the delays would indicate the likelihood of a source in the near field, but this eventuality is not yet solved for.  The high time-resolution correlation is then repeated with the refined position for the transient source, and the calibration and fringe-fitting steps are repeated to ensure that an accurate location has been derived.  The entire process is repeated if significant delays remain.  Once a sufficiently accurate position has been determined, the data are calibrated and averaged in frequency.  Given a delay precision of order tens of nanoseconds (which is achievable at our detection S/N for continuum bandwidths; higher S/N events would give more accurate delays) we can expect source localization with an accuracy of no worse than an arcsecond.
 
The visibilities can now be imaged, maintaining fine time resolution in order to be able to measure on-pulse power and off-pulse background sky noise. If external phase reference solutions were available from calibrator data and utilized in preference to the FRING-derived solutions for the transient source, the uncertainty in the reference frame will be very small, allowing very precise localization of the transient source (which will appear somewhere within an arcsecond of the image center).  If the FRING-derived solutions were used, this effective``self--calibration'' will mean that the source will appear exactly at the image phase center, but the absolute position will be uncertain by 0.1 -- 1 arcseconds, due to the uncertainty in the delay solutions used to derive the source offset.

\section{Test Data and Results}
Initial development and testing of the V-FASTR system was undertaken under a pilot program approved as VLBA project BT100.
In order to test the final full system, dedicated observations of pulsars B0329+54 and B0531+21 (the Crab pulsar) were taken
under VLBA program BM348. Pulsar B0329+54 is bright enough that most individual pulses can be seen in the raw data,
whereas for the Crab pulsar we do not expect to see individual pulses but should see giant pulses. DMs and
other useful information for the pulsars was used from the ATNF pulsar database \citep{2005AJ....129.1993M}.\footnote{\url{http://www.atnf.csiro.au/research/pulsar/psrcat/}}

The observation setup used four non-contiguous frequency bands centered on 1410~MHz, 1504~MHz, 1590~MHz and 1680~MHz, each with 8~MHz bandwidth and two polarizations.
During correlation, spectrometer data were generated with 1.6~ms time resolution and 0.25~MHz frequency
resolution. For B0329+54, seven sets of 240~s scans were performed interspersed between calibrator scans.
For the Crab, six sets of 500~s scans were performed between calibrator scans.
One of the VLBA antennas was not available during the observation so a total of nine was used in BM348.

\begin{figure*}
 \centering
 \includegraphics*[scale=0.55]{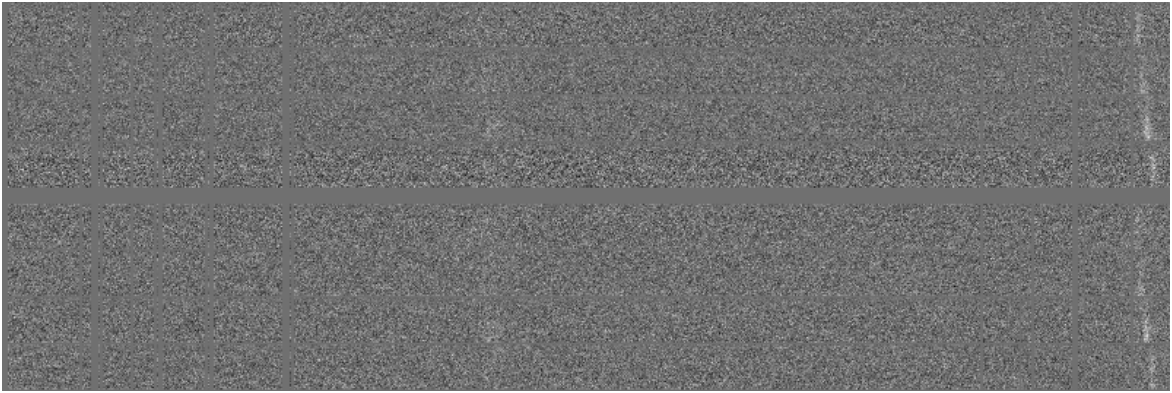}
 \caption{Example dynamic spectrum from pulsar B0329+54. Time runs left to right and each pixel is 1.6~ms. Non-contiguous
        frequency bands are stacked vertically and grouped into the two polarizations. A relatively strong pulse is seen to
        the far right. A weaker pulse is left of center. Note the discontinuities in time over the dispersed pulse due to
        non-contiguous frequency bands.
        }
 \label{fig:B0329_raw_pulse}
\end{figure*}

\begin{figure}
 \includegraphics*[scale=0.5]{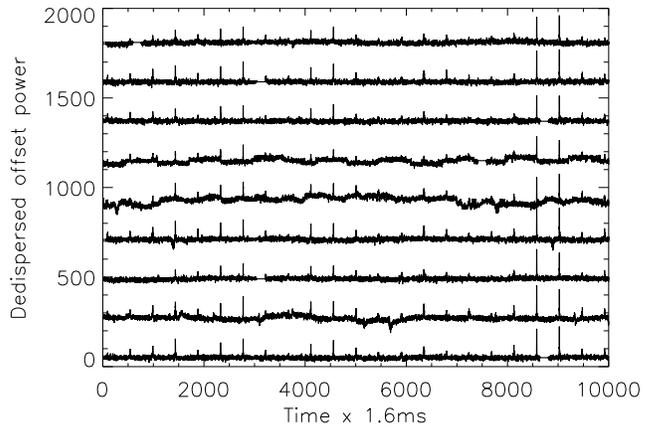}
 \caption{Dedispersed power time series from all VLBA antennas for 10,000 time steps (16 s) of data. Successive antennas
          are offset in power for clarity but actually all data streams have zero mean.}
 \label{fig:B0329_all_streams}
\end{figure}

Figure \ref{fig:B0329_raw_pulse} shows an example dynamic spectrum from pulsar B0329+54 for a single VLBA antenna.
A relatively strong dispersed pulse is seen clearly at the far right and a weaker pulse is left of center.
Figure~\ref{fig:B0329_all_streams}
shows the dedispersed power time series for B0329+54 for all nine of the antennas that took part in the observation.
The plots have been offset in power for clarity.
Pulses from the pulsar are clearly seen varying in strength with time, but the power from any individual pulse correlates well between the antennas.

Because most individual pulses from B0329+54 can be seen in the dedispersed data and the pulsar's timing properties are already well known, the accuracy of the detection system can be thoroughly evaluated by comparing candidate transient events with known pulse arrival times and off-pulse times. A detailed analysis of the detection system performance using B0329+54 appears in a companion paper \citep{Thompson2010}.

Also noteworthy is the effect of RFI, which causes the dedispersed median-subtracted power to systematically wander away from zero in both the positive and negative directions over time for some antennas. This is most evident in the fifth and sixth data streams, counting from the bottom.


\begin{figure}
 \includegraphics*[scale=0.5,angle=270]{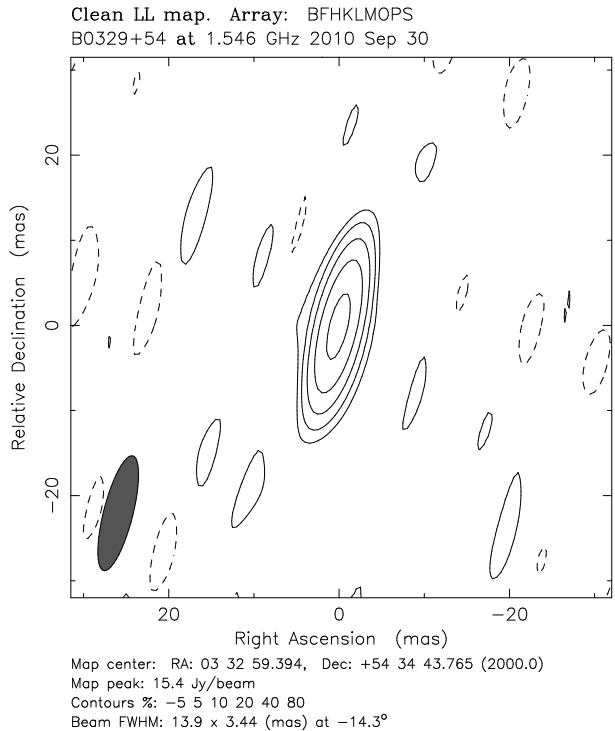}
 \caption{Image of a single pulse from B0329+54. The synthesized beam is shown in gray at lower left. Contours are relative to the peak, with the -5\% contour dashed.}
 \label{fig:B0329_pulse_image}
\end{figure}

To test the post-detection phase of the data processing, a snippet of raw data containing a single pulse from B0329+54 was used to
form an image of a pulse. With the exception of determining the phase center, the follow-up strategy described in Section~\ref{sec:follow-up} was employed.
A 4~ms section of on-pulse data was used to form the image. The resulting pulse is shown in Figure \ref{fig:B0329_pulse_image}. The pulse is clearly seen with an S/N of approximately 50:1.

The second set of data from VLBA program BM348 totaled approximately 50 minutes on the Crab pulsar.
Normal pulses from the Crab pulsar are not strong enough that individual pulses can be seen in the dedispersed time series, however
the Crab emits giant pulses intermittently. Giant pulses are exactly the kind of transient (non-periodic) signal that V-FASTR aims to detect, hence the point of these observations was to test the system with realistic data that are likely to contain a transient signal.

\begin{figure}
 \includegraphics*[scale=0.5]{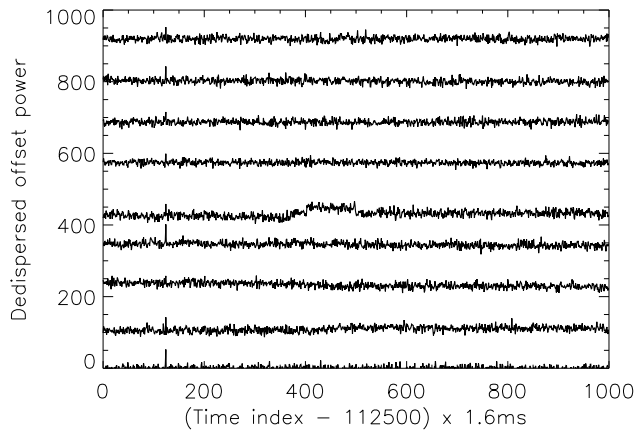}
 \caption{Dedispersed power time series from all VLBA antennas around a giant pulse from the Crab pulsar.
         }
 \label{fig:Crab_all_streams}
\end{figure}

A single convincing giant pulse event was seen from the Crab pulsar in our data. Figure~\ref{fig:Crab_all_streams} shows the dedispersed time series for all antennas, again offset in power for clarity.

Based on previous observations of Crab giant pulses, we aim to estimate roughly how many events we should have seen.
Emission from the Crab nebula raises the antenna temperature, making
the $1\sigma$ sensitivity of each antenna approximately 6.7~Jy over 1~ms. Interstellar scintillation affects the observed flux density of the pulses \citep[e.g.,][]{1995ApJ...453..433L}, however based on \citet{2008ApJ...676.1200B} we expect to see roughly a single 10~kJy pulse over one hour with $1\mu$s time resolution at 1.4~GHz. Our data have 1.6~ms time resolution so a 10~kJy/1~$\mu$s pulse would have an average flux density of $\sim10$~Jy in our data. Hence, such a pulse would be a $\sim2\sigma$ detection in any one antenna. Therefore, it appears quite reasonable that only a single convincing event was seen given the time resolution of the data.

\section{Discussion}

We have described V-FASTR, a commensal experiment to search for fast transient radio signals in real time during the normal correlation processing of a radio interferometer array, the VLBA. V-FASTR is designed for multiple-antenna arrays that are normally used for interferometry, so includes novel techniques to combine the signals from different antennas that maintains the expected sensitivity of the incoherently combined array, but is highly robust to RFI in any antenna.

The V-FASTR experiment is different from past and ongoing transient searches in that it makes use of a VLBI array.  This imposes some additional challenges, such as the significant delay, typically one week, between observing and detection, possibly compromising follow-up observations. However, there are several advantages to this approach that make the V-FASTR approach a very compelling transient search strategy.
Because the processing pipeline uses incoherent dedispersion and detection,
the experiment is sensitive to the entire primary beam of the VLBA antennas, which is an advantage over experiments that use
large single dishes due to the increased FOV of the (relatively) smaller 25~m VLBA antennas ($\sim 0.5^{\circ} $ at 1.4~GHz).
The reduction in sensitivity compared to a large single dish is offset by the addition of signals from all antennas in the array.
Also, the disadvantage of the delay between observation and correlation may be removed in the future, with the adoption of \emph{e}-VLBI (i.e., real-time streaming of data from telescopes to the correlator) by the international VLBI community.

\begin{table}
\begin{center}
\caption{Comparison of fast transient experiments.
\label{tab:comparison}}
\begin{tabular}{ccccc}
\tableline
\tableline
Experi- & Instru- & FOV       & Sensitivity & Duration \\
ment    & ment    & deg$^{2}$ & (Jy)        & (hr) \\
\tableline
FE\tablenotemark{a} & ATA & 150 & 20 & 120 \\
B10\tablenotemark{b} & Parkes & 0.53\tablenotemark{*} & 0.05 & 1078 \\
L07\tablenotemark{c} & Parkes & 0.53\tablenotemark{*} & 0.05 & 481 \\
HTRU\tablenotemark{d} & Parkes & 0.53\tablenotemark{*} & 0.05 & Ongoing \\
PALFA\tablenotemark{e} & Arecibo & 0.17\tablenotemark{@} & 0.01 & 548 + Ongoing \\
V-FASTR & VLBA & 0.27 & 0.3 & 4400\tablenotemark{v}
\end{tabular}
\tablenotetext{a}{ATA Fly's Eye \citep{2011AAS...21724006S} and priv. comm.}
\tablenotetext{b}{\citet{2011ApJ...727...18B}.}
\tablenotetext{c}{\citet{2007Sci...318..777L}.}
\tablenotetext{d}{\citet{2010MNRAS.409..619K}.}
\tablenotetext{e}{\citet{2009ApJ...703.2259D}.}
\tablenotetext{*}{Using the 13 beam multibeam receiver on the Parkes telescope.}
\tablenotetext{@}{Using the 7 beam multibeam receiver on the Arecibo telescope.}
\tablenotetext{v}{Assuming six months of usable on-sky time over a year.}
\end{center}
\end{table}

Table \ref{tab:comparison} shows a brief summary comparing V-FASTR to previous and current experiments that include searches for single pulses where
FE is the ATA ``Fly's Eye'' system \citep{2011AAS...21724006S},
B10 is for the archival search of \citet{2011ApJ...727...18B},
L07 is for the archival search of \citet{2007Sci...318..777L},
HTRU is the ongoing pulsar/transient survey of \citet{2010MNRAS.409..619K},
and PALFA is the ongoing pulsar/transient survey of \citet{2009ApJ...703.2259D}.
The most relevant performance metrics for this kind of experiment, where the properties of the source population are not known, are the FOV, the sensitivity and the time spent observing.
In the table, the FOV and sensitivity columns assume an observing frequency of 1.4~GHz.
The sensitivity ($1\sigma$) has been calculated simply using the radiometer equation where the integration time has been normalized to
1~ms time samples, but the actual bandwidth of the instrument has been retained. It makes no corrections for the loss in sensitivity
due to, for instance, the data being quantized to 1 bit or the frequency resolution of the spectrometer channels.
The Fly's Eye is sensitive to spatially rare, strong bursts whereas the two Parkes-based experiments are sensitive to more frequent, fainter
bursts. After a few months of observing time, V-FASTR should roughly equal the sensitivity of the Parkes experiments
(in terms of time on sky) and have the additional benefits of using an interferometer array.

A significant strength of the V-FASTR search is its potential to localize, and perhaps image at milliarcsecond resolution, the transient emission. Baseband data from each station are available at the time of detection.  Small ($\sim 1$~s duration) segments of these baseband data can be extracted for suitably interesting transient candidates.  These data can be reprocessed with significantly higher time resolution and even coherently dedispersed if necessary to further characterize the parameters of the detection.
High time and spectral resolution correlation products can be generated from this extracted baseband data.  Calibration data, which in most cases are generated as the main product from the DiFX software correlator, can be used to calibrate and image the transient.

Another important advantage V-FASTR has over other experiments is its geographic extent.  The occurrence of false positives can be drastically reduced by requiring simultaneous detection at multiple antennas. Man-made signals from ground-, aircraft-, or spacecraft-based transmitters can be ruled out either by requirements of mutual visibility or simultaneity.  For example, signals received from a geostationary satellite at an orbital radius of 42,000~km would be received with a 0.7~ms timing curvature across the VLBA, within the detection limit of this experiment; this signal curvature filter becomes increasingly effective at lower altitudes.

Many aspects of the V-FASTR design and implementation are relevant to the fast transient science programs of next generation radio arrays that are currently under construction, such as ASKAP's CRAFT fast transient project \citep{2010PASA...27..272M,2010CRAFT}. Indeed, V-FASTR is considered a ``trailblazer'' project for CRAFT, aimed at highlighting issues and finding new ways to approach the challenges, as well as being a significant science project in its own right.
The trailblazing aspect of V-FASTR is already bearing fruit. For example, the analysis of robust detection algorithms discussed in the companion paper \citep{Thompson2010} show that it is highly advantageous to dedisperse data streams from each telescope separately and combine them later in a robust detector.
As a guide to arrays with large numbers of antennas where it may not be feasible to dedisperse the data from all antennas separately, it may still be advantageous to group antennas into sub-groups where the data streams from all antennas within a subgroup are combined before dedispersion, but different subgroups are combined after dedispersion in a robust detector.

V-FASTR provides a ready-made framework for similar commensal fast transient projects at other institutes using the DiFX correlator.  Regular DiFX users include VLBI operations such as the Australian Long Baseline Array (LBA; $\sim$4 weeks year$^{-1}$ observing) and the Max Planck Institute for Radio Astronomy correlator facility (MPIfR; near full time operation covering geodetic and astronomical experiments).  A number of other facilities use DiFX on a trial basis and may become available for transient operations in the future.  Finally, several other facilities employ different software correlators which would be amenable to fast transient projects utilizing a similar framework to V-FASTR; these include LOFAR and the Giant Metrewave Radio Telescope (GMRT), both of which have considerably higher collecting area and hence sensitivity compared to the VLBA, and which operate near full time at relatively low frequencies.

The VLBA, through its course of science-driven observing, works over a wide frequency range (330~MHz to 90~GHz).  Further, the VLBA antennas are typically pointed at astronomical objects of potential interest, including, but not limited to, active stars, active galaxies, the Galactic center and pulsars.
To determine the type of data that are processed at the VLBA correlator the following statistics were generated for all scientific observations correlated by the VLBA correlator in the months of 2010 October, November and December.  Thirty-eight percent of projects aimed to study Galactic sources with the remainder observing extragalactic sources.  Since astronomical calibration is typically derived from observations of extragalactic sources, it is expected that only about 25\% of the time is actually spent targeting Galactic objects with the vast majority of the remaining 75\% spent looking at the nuclei of galaxies. Observing time was split between the various observing bands as follows: 8\% in 1.2--1.8~GHz, 13\% in 2--2.5~GHz, 1\% in 4.5--5~GHz, 24\% in 8--9~GHz, 19\% in 14--16~GHz, 23\% in 21--24~GHz, 10\% in 40--50~GHz and 3\% in 80--90~GHz.  About 25\% of observations targeted specific spectral lines.  However, in most of cases the observations were set up to observe with wide bandwidth for the purpose of calibration.  Note that the observing statistics vary greatly month to month in conjunction with patterns of favorable climate and other factors.
The V-FASTR experiment is therefore a ``blind'' search for transient signals, but is not unbiased because the telescopes are not pointed at random locations in the sky.

Finally, the V-FASTR project offers the potential to supplement existing performance monitoring capabilities of the VLBA antennas and correlator.  RFI monitoring on timescales shorter than one correlator integration period can be investigated for the first time with the spectral kurtosis output being a possible new source of information that astronomers could use to identify and excise portions of time/frequency that may be subtly contaminated.  The detection of the switched noise calibration signal could provide, for the first time to the VLBA, amplitude calibration data that are determined as a function of frequency within a single baseband channel.  This capability will be increasingly attractive as the bandwidth per baseband channel increases in conjunction with the ongoing VLBA bandwidth expansion project.

V-FASTR has been approved for one year of commensal operation under VLBA project code BT111. Within that period we hope to obtain at least
six months of data at a variety of frequencies. As noted in Table \ref{tab:comparison}, V-FASTR will be comparable to the large searches
based on Parkes telescope archival data with approximately one month of on-sky time. We therefore expect that the experiment will provide
new and significant insight into the population of transient radio sources.

\acknowledgments

The International Centre for Radio Astronomy Research is a Joint Venture between Curtin University and The University of Western Australia, funded by the State Government of Western Australia and the Joint Venture partners.
S.J.T is a Western Australian Premier's Research Fellow.
R.B.W is supported via the Western Australian Centre of Excellence in Radio Astronomy Science and Engineering.
A.T.D is supported by an NRAO Jansky Fellowship.
Part of this research was carried out at the Jet
Propulsion Laboratory, California Institute of Technology,
under contract with the US National Aeronautics and Space
Administration.
The National Radio Astronomy Observatory is a facility of the National Science Foundation operated under cooperative agreement by Associated Universities, Inc.
AIPS is produced and maintained by the National Radio Astronomy
Observatory, a facility of the National Science Foundation
operated under cooperative agreement by Associated Universities, Inc.

{\it Facility:} \facility{VLBA}

\bibliographystyle{apj}
\bibliography{vfastr_refs}

\clearpage

\end{document}